\documentclass[twocolumn]{aastex6}

\usepackage{natbib,graphicx,wasysym}
\graphicspath{{./figs/}}

\fullcollaborationName{}

\begin{document}

\title{An Optical and Infrared Time-Domain Study of the Supergiant Fast X-ray Transient Candidate IC 10 X-2}

%% Optical and infrared time variations of IC 10 X-2, a 

%% Use \author, \affil, plus the \and command to format author and affiliation 
%% information.  If done correctly the peer review system will be able to
%% automatically put the author and affiliation information from the manuscript
%% and save the corresponding author the trouble of entering it by hand.
%%
%% The \affil should be used to document primary affiliations and the
%% \altaffil should be used for secondary affiliations, titles, or email.

%% Authors with the same affiliation can be grouped in a single
\author{Stephanie Kwan\altaffilmark{1}, Ryan M. Lau\altaffilmark{1}, Jacob Jencson\altaffilmark{1}, Mansi M. Kasliwal\altaffilmark{1}, Martha L. Boyer\altaffilmark{2,3}, Eran Ofek\altaffilmark{4}, Frank Masci\altaffilmark{5}, Russ Laher\altaffilmark{5}}

\altaffiltext{1}{Division of Physics, Mathematics and Astronomy, California Institute of Technology (Caltech), 1200 E. California Blvd., Pasadena, CA 91125 USA}
\altaffiltext{2}{Observational Cosmology Lab, Code 665, NASA Goddard Space Flight Center, Greenbelt, MD 20771 USA}
\altaffiltext{3}{Oak Ridge Associated Universities (ORAU), Oak Ridge, TN 37831 USA}
\altaffiltext{4}{Department of Particle Physics and Astrophysics, Weizmann Institute of Science, Rehovot 76100, Israel}
\altaffiltext{5}{Infrared Processing and Analysis Center, Caltech, 1200 E. California Blvd., Pasadena, CA  91125 USA}

\begin{abstract}
We present an optical and infrared (IR) study of IC 10 X-2, a high-mass X-ray binary in the galaxy IC 10. Previous optical and X-ray studies suggest X-2 is a Supergiant Fast X-ray Transient: a large-amplitude (factor of $\sim$ 100), short-duration (hours to weeks) X-ray outburst on 2010 May 21. We analyze R- and g-band light curves of X-2 from the intermediate Palomar Transient Factory taken between 2013 July 15 and 2017 Feb 14 show high-amplitude ($\gtrsim$ 1 mag), short-duration ($\lesssim8$ d) flares and dips ($\gtrsim$ 0.5 mag). Near-IR spectroscopy of X-2 from Palomar/TripleSpec show He I, Paschen-$\gamma$, and Paschen-$\beta$ emission lines with similar shapes and amplitudes as those of luminous blue variables (LBVs) and LBV candidates (LBVc). Mid-IR colors and magnitudes from \textit{Spitzer}/IRAC photometry of X-2 resemble those of known LBV/LBVcs. We suggest that the stellar companion in X-2 is an LBV/LBVc and discuss possible origins of the optical flares. Dips in the optical light curve are indicative of eclipses from optically thick clumps formed in the winds of the stellar counterpart. Given the constraints on the flare duration ($0.02 - 0.8$ d) and the time between flares ($15.1\pm7.8$ d), we estimate the clump volume filling factor in the stellar winds, $f_V$, to be $0.01 < f_V < 0.71$, which overlaps with values measured from massive star winds. In X-2, we interpret the origin of the optical flares as the accretion of clumps formed in the winds of an LBV/LBVc onto the compact object.
\end{abstract}

%We report the existence and characteristics of a mid-infrared (IR) counterpart to IC 10 X-2, a compact object-supergiant binary in the galaxy IC 10. Previous optical and X-ray studies argue that the system resembles a Supergiant Fast X-ray Transient: a large-amplitude (factor of $\sim$ 100), short-duration (hours to weeks) X-ray outburst on 2010 May 21. We track its mid-IR variability over ten epochs of archival imaging data from the \textit{Spitzer Space Telescope} throughout 2004-2016 and present \textit{JHK}$_S$ imaging and spectroscopy taken with the 200-inch Hale Telescope at Palomar Observatory in July 2016. \textbf{We find evidence of mid-IR variability in the 5 months preceding the X-ray outburst.} \textbf{Based on the near-IR spectrum, mid-IR color, and luminosity of IC 10 X-2, we infer that the stellar companion in this HMXB system is a luminous blue variable (LBV) star, making this one of the first candidates for an LBV-HMXB. We then speculate on the origin of the fast X-ray outburst by evaluating wind-fed accretion scenarios for SFXT outbursts assuming the mass donor star is an LBV.

%Our analysis suggests that the IC 10 X-2 X-ray luminosities and variability timescales are consistent with the interpretation where the fast X-ray outburst occurs due to the motion of the compact object through an asymmetric outflow from the LBV companion.


\section{Introduction} \label{sec:intro}
%\subsection{X-ray emitting binary systems}

High-mass X-ray binaries (HMXBs) are accretion-powered binary systems consisting of a neutron star (NS) or black hole accreting matter from its companion high-mass star ($> 10 M_{\astrosun}$). HMXBs are historically divided into two classes based on their method of accretion: BeXBs, which accrete from circumstellar disks around an emission-line B (Be) donor star, and supergiant HMXBs (sgHMXBs), which have Roche lobe overflow or stellar winds from an early spectral type supergiant OB donor stars (see \citealt{Chaty2014} for a review).

Since its launch in 2002, \textit{INTEGRAL} has led to the discovery of a subclass of sgHMXBs called supergiant fast X-ray transients (SFXTs). SFXTs exhibit unusually high-amplitude ($\sim$ 1000), bright ($\sim10^{36}$-10$^{37}$ erg s$^{-1}$), fast (minutes to days) outbursts compared to those of regular sgHMXB \citep{Ducci}. These outbursts are inconsistent with orbital motion of the neutron star through a smooth medium. Several accretion models have been proposed. One group of models attributes the outbursts with accretion of denser ``clumps'' in a inhomogeneous, anisotropic stellar wind (e.g. \citealt{Sidoli, Walter, Negueruela}). Another group involves accretion-gating mechanisms, where variations in accretion onto the compact object arise from transition stages between magnetic and centrifugal barriers (see \citealt{Illarionov, Bozzo, Drave} and references within). \citet{Liu} suggests that some wind-fed sgHMXBs evolved from BeHMXBs, but the relationship between SFXTs and sgHMXBs remains unclear. 

\citet{Laycock} reported the discovery of IC 10 X-2, another HMXB, based on Chandra measurements of X-ray outbursts in 2003 and 2010. The aspect corrected coordinates for IC 10 X-2 are RA: $00^h 20^m 20^s.94$, Dec: $59^\circ 17^m 59^s.0$, with a 95\% confidence region of radius 0.6''. It was reported to exhibit an X-ray outburst with peak luminosity $1.8 \times 10^{37}$erg s$^{-1}$ for the best-fitting power-law spectral model. %This peak X-ray luminosity was comparable (an order of magnitude larger) than that of (SN) 2010da. 
A second outburst in 2010 May (MJD 55337.5) was constrained to have a duration less than 3 months. Based on IC 10 X-2's X-ray and optical spectra, \citet{Laycock} posited that its primary star is a nitrogen-enriched B supergiant star, calling it a ``luminous blue supergiant''. The B-type supergiant appearance opens the question of whether the primary star is a luminous blue variable (LBV).

%\subsection{Luminous Blue Variables}
LBVs are massive evolved stars that exhibit extreme luminosities ($\sim10^5 -10^6$ L$_\odot$), strong spectral variability ($T_\mathrm{eff}\sim7000 - 20000$ K), and eruptive mass loss as high as $\sim1$ M$_\odot$ yr$^{-1}$ on decade-long timescales~\citep{Smith}. LBVs, however, are difficult to classify due to the range of spectroscopic and photometric properties that they span. %LBVs show spectroscopic and photometric variability on timescales of years to decades due to changes in the stars' apparent temperature and radius \citep{Vink}. %They also exhibit smaller amplitude variability on shorter timescales of weeks and months, though this is a feature of supergiants in general.

\begin{figure*}[t!]
	\plotone{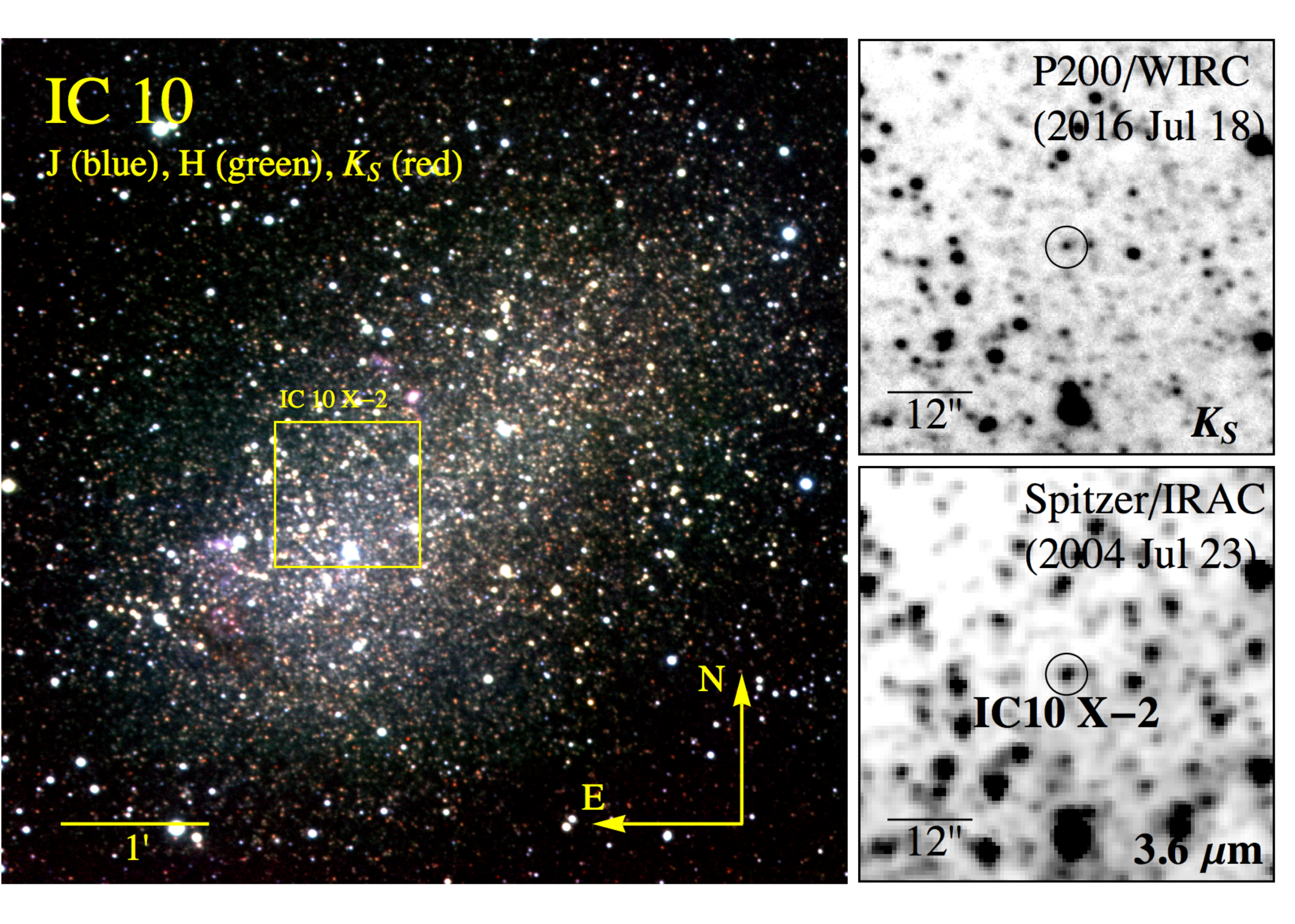}
        \caption{False color image of IC 10 taken with J (blue), H (green), and K$_S$ (red) band filters by P200/WIRC on 2016 July 18. The  $1\times1$' yellow box overlaid on the image is centered on IC 10 X-2 and corresponds to the field of view of the P200/WIRC K$_S$ and \textit{Spitzer}/IRAC Ch1 images shown on the right. The black circle overlaid on the images in the right is centered on the near- and mid-IR counterparts of IC 10 X-2 and represents the 3''-radius aperture used to perform the photometry.}
        \label{fig:IC10X2Color}
\end{figure*}

%LBVs are difficult to classify due to the range of spectroscopic and photometric properties that they span. %
%Several of LBVs' unpredictable outbursts can mimic supernova (SN) and are dubbed supernova ``impostors'', since they resemble SNe IIn spectroscopically but are typically fainter and are non-terminal explosions where the progenitor star survives \citep{Ofek, Thone}. One such example is impostor supernova (SN) 2010da, which was discovered as an optical transient in the nearby galaxy NGC 300 on 2010 May 23 \citep{Monard}. Initially classified as a supernova, it was next thought to be a HMXB with a LBV progenitor, based on mid-IR photometry from archival \textit{Spitzer} imaging data and optical spectroscopy. However, \citet{Lau} recently argued that the progenitor star is not a LBV, but instead a more common supergiant B-type star (sgB[e]), based on its luminosity and dust mass comparisons with known sgB[e] stars in the Large Magellanic Cloud. 

%\textbf{Mid-IR observations have indeed been demonstrated as useful diagnostics of the stellar counterpart and circumstellar environment of luminous X-ray systems (e.g. \citealt{Lau2017}).}

%\subsection{IC 10 X-2: a HMXB with a LBV companion?}

In order to investigate the nature of IC 10 X-2, we study its optical and IR counterpart using data from the intermediate Palomar Transient Factory (iPTF) and archival multi-epoch imaging observations taken by  \textit{Spitzer/IRAC}. In addition, we present $JHK_s$ photometry and near-IR spectroscopy of IC 10 X-2 taken at Palomar Observatory in 2016 July. We compare the near-IR spectroscopic features and the mid-IR color/magnitude of X-2 against various classes of supergiant stars to classify the stellar counterpart. We then study the optical light curve and consider possible accretion scenarios that result in SFXT flares.

\floattable
\begin{deluxetable}{ccClc}[t!]
\tablecaption{IC 10 X-2 IR fluxes\label{tab:FluxLog}}
\tablecolumns{5}
\tablewidth{0pt}
\tablehead{ \colhead{MJD} & \colhead{UT start time} &
\colhead{Filter} & \colhead{Fluxes\tablenotemark{a}} 
& \colhead{Inst. or Database}\\
\colhead{} & \colhead{} & \colhead{(s)} & \colhead{(mJy)} & \colhead{} }
\startdata
2000-09-16 & 51803 & $JHK_s$ & 0.419 $\pm$ 0.06, 0.708*, 0.421* & 2MASS  \\
2004-07-23 & 53209 & [3.6], [4.5], [5.8], [8.0] & 0.335 $\pm$ 0.02, 0.282 $\pm$ 0.01, 0.21 $\pm$ 0.04, 0.20 $\pm$ 0.03 & IRAC \\
2010-01-29 & 55225 & [3.6], [4.5] & 0.236 $\pm$ 0.02,  0.184 $\pm$ 0.01 & IRAC\\
2010-02-19  & 55246 & [3.6], [4.5] & 0.196 $\pm$ 0.02, 0.141 $\pm$ 0.01 &IRAC \\
2010-03-10 & 55265 & [3.6], [4.5] & 0.234 $\pm$ 0.02, 0.152 $\pm$ 0.01 & IRAC\\
2010-09-09 & 55448 & [3.6], [4.5] & 0.216 $\pm$ 0.02, 0.159 $\pm$ 0.01 & IRAC\\
2010-10-04 & 55473 & [3.6], [4.5] & 0.223 $\pm$ 0.02, 0.164 $\pm$ 0.01 &IRAC\\
2010-10-14 & 55483 & [3.6], [4.5] & 0.198 $\pm$ 0.02, 0.146 $\pm$ 0.01 &IRAC \\
2011-09-24 & 55828 & [3.6], [4.5] & 0.199 $\pm$ 0.02, 0.150 $\pm$ 0.01  & IRAC (DUSTiNGS\tablenotemark{b})\\
2012-04-04 & 56021 & [3.6], [4.5] & 0.214 $\pm$ 0.02, 0.152 $\pm$ 0.01 & IRAC (DUSTiNGS\tablenotemark{b}) \\
2015-03-11 & 57092 & [3.6], [4.5] & 0.184 $\pm$ 0.02, 0.138 $\pm$ 0.01 & IRAC (DUSTiNGs\tablenotemark{c}) \\
2016-03-23 & 57470 & [3.6], [4.5] & 0.199 $\pm$ 0.02, 0.145 $\pm$ 0.01 & IRAC (DUSTiNGs\tablenotemark{c})\\
2016-07-18 & 57587 & $JHK_s$ & 0.382 $\pm$ 0.06, 0.346 $\pm$ 0.03, 0.396 $\pm$ 0.08 & WIRC\tablenotemark{d}\\
\enddata
\tablenotetext{a}{* denotes upper limits.}
\tablenotetext{b}{PID 80063, \citet{Goldman}.}
\tablenotetext{c}{PID 11041, \citet{Goldman}.}
\tablenotetext{d}{The WIRC instrument was dithered at five positions in three quadrants (avoiding a quadrant with bad pixels).}
%\tablecomments{The WIRC instrument was dithered at five positions in three quadrants (avoiding a quadrant with bad pixels).}
\end{deluxetable}

\section{Observations and Data Reduction} \label{sec:observationsAnalysis}
The primary source of our near-IR data was the Infrared Array Camera (IRAC) onboard the \textit{Spitzer} Space Telescope. In its cold phase, IRAC imaged in four channels denoted by their wavelengths in microns ([3.6], [4.5], [5.8], and [8.0]). On 2009 May 15, IRAC transitioned to its warm phase, imaging only in channels [3.6] and [4.5]. X-2 was observed once during the cold phase and ten times during the warm phase. 

X-2 was also imaged by the DUSTiNGS (Dust in Nearby Galaxies with Spitzer) program, a sample of 50 dwarf galaxies within 1.5 Mpc mapped with IRAC channels [3.6] and [4.5] \citep{Boyer}. 

In addition, we performed near-IR imaging in the $JHK_s$ channels and near-IR spectroscopy at Palomar Observatory on 2016 Jul 18. All IR fluxes are listed in  Table \ref{tab:FluxLog}.

\subsection{Spitzer photometry} \label{subsec:SpitzerPhotometry}
We demonstrate mid-IR variability concurrent with the 2010 X-ray outburst by constructing a light curve using archival \textit{Spitzer} IRAC images. The flux density was calculated using a circle with 2.5'' radius centered at X-2. Background flux density was calculated from an annulus centered at IC 10 X-2 with 3.5'' inner radius and 6'' outer radius to avoid contamination from nearby bright sources. The raw data (counts) were photometrically calibrated using IRAC specifications from \citet{2MASS}, giving fluxes in mJy. The 1-$\sigma$ uncertainty in the counts and fluxes were calculated by multiplying the uncertainty in the mean counts of the background annulus by the area of the 3'' circle. 

\subsection{P200 Near-IR Imaging and Spectroscopy} \label{subsec:WIRC}
To complement the \textit{Spitzer} images, we obtained near-infrared images of IC 10-X2 in the $J$, $H$, and $K_s$ bands on 2016 July 18 using the Wide Field Infrared Camera (WIRC; \citealt{Wilson}) on the 200-in. Hale Telescope at Palomar Observatory (P200).  To allow accurate subtraction of the sky background, we took well-dithered sequences of 18 exposures of 60 s integration in $J$, 19 exposures of 25 s integration with 2 coadds in $H$, and 22 exposures of 15 s integration with 2 coadds in $K_s$, for total integration times of 1080 s in $J$, 950 s in $H$, and 660 s in $K_s$.  Imaging reductions, including flat-fielding, background subtraction, astrometric alignment, and stacking of individual frames were performed using a custom pipeline. The photometric zero points of the final images were measured using aperture photometry of 7 isolated 2MASS stars in the field. Flux density was calculated using a circle with 3'' radius centered at IC 10 X-2. Background flux density was calculated from an annulus centered at IC X-2 with 5'' inner radius and 10'' outer radius. Fluxes were converted to magnitudes assuming a Gaussian distribution of combined uncertainties from the 2MASS star zero-points and the X-2 photometry.

%Add aperture size and clarify the following statement about incorporating both uncertainties
%\subsection{P200 Near-IR Spectroscopy} \label{subsec:Spectroscopy}
Near-infrared spectroscopy was performed using the Triple Spectrograph (TripleSpec) instrument during the same observation run to complement visible-spectrum spectroscopy by \citet{Laycock}. We observed IC 10-X2 on 2017 July 8 with the near-IR Triple Spectrograph (TripleSpec) on the 200-in. Hale Telescope at Palomar Observatory. TripleSpec uses a $30 \times 1$~arcsec slit to obtain a $JHK$ spectrum from 1--2.4~$\mu$m with a resolving power of 2500--2700. We obtained a sequence of 8 300~s exposures, for a total integration time of 2400 s. TripleSpec data were reduced using standard procedures with bias and flat field corrections. Wavelengths were calibrated against atmospheric lines.

%Reduction was performed by R. Lau using the Modest Image Analysis and Reduction (MIRA) software written by Terry Herter for mid-IR imaging data.

\subsection{iPTF g- and R-band Photometry}
We used optical photometry of IC 10 X-2 from the IPAC/iPTF Discovery Engine (or IDE), which is described by \citet{Masci}. Raw images from the Palomar 48-inch Samuel Oschin Schmidt Telescope and are run through a real-time pipeline, which performs basic calibration, then sent to PTFIDE, the image-differencing and transient extraction module. The images are returned to the pipeline for archiving, database-loading, and machine-learned vetting. The overall products of the iPTF pipeline and PTFIDE are difference images, mask and uncertainty images, and QA metrics, in addition to candidate transient catalogs with PSF-fit and aperture photometry, and thumbnail images of transient candidates. Optical light curves in the R and g bands, spanning 2010 Nov 14 to 2017 Feb 14 and 2009 Aug 25 to 2017 Jan 10 respectively, were generated for IC 10 X-2. 

\section{Results and Analysis} \label{sec:Results}
\subsection{IR and Optical Light Curves of IC 10 X-2}  \label{subsec:lc}
\begin{figure*}[ht]
	\plotone{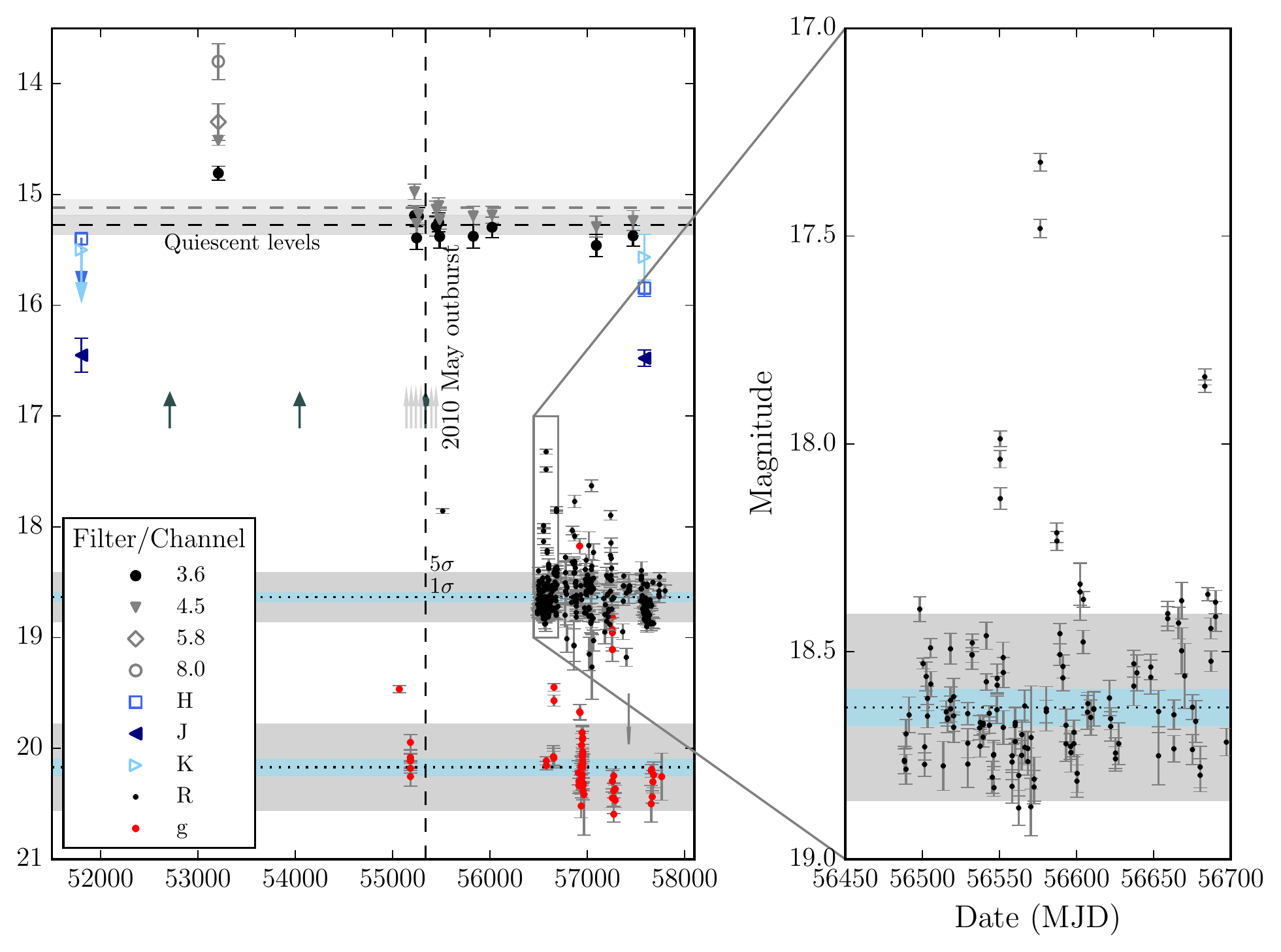}
	\caption{IC 10 X-2 near-IR and optical light curve, constructed from 11 \textit{Spitzer} epochs (spanning 2004 Jul 23/52309 MJD to 2016 Mar 23/57470 MJD), 1 WIRC epoch, and optical data from iPTF. The [3.6] and [4.6] flux densities (mJy) drop by factors of 0.30 ($\sim 2\sigma$) and 0.35 ($\sim 4\sigma$) respectively shortly before the 2010 May X-ray outburst (dashed line, 55337 MJD). The quiescent levels for [3.6] and [4.5] were calculated as a simple average of the four most recent fluxes.
    \newline The arrows on the first set of $H$ and $K_s$ indicate that those fluxes are upper limits from the 2MASS database (size of errors unknown). The upward-pointing dark grey arrows indicate the X-ray detection epochs from \citet{Laycock}; the upward-pointing light gray arrows indicate non-detections. 
    \newline In the optical bands, the median magnitude is plotted, with 1 $\sigma$ and 5 $\sigma$ levels denoted.
    \newline Inset: A close-up of a densely sampled R-band light curve, showing several flares (defined to be $\geq 5 \sigma$ detections).
    }
	\label{fig:IR_and_optical_lc}
\end{figure*}

\begin{figure*}[ht]
	\plotone{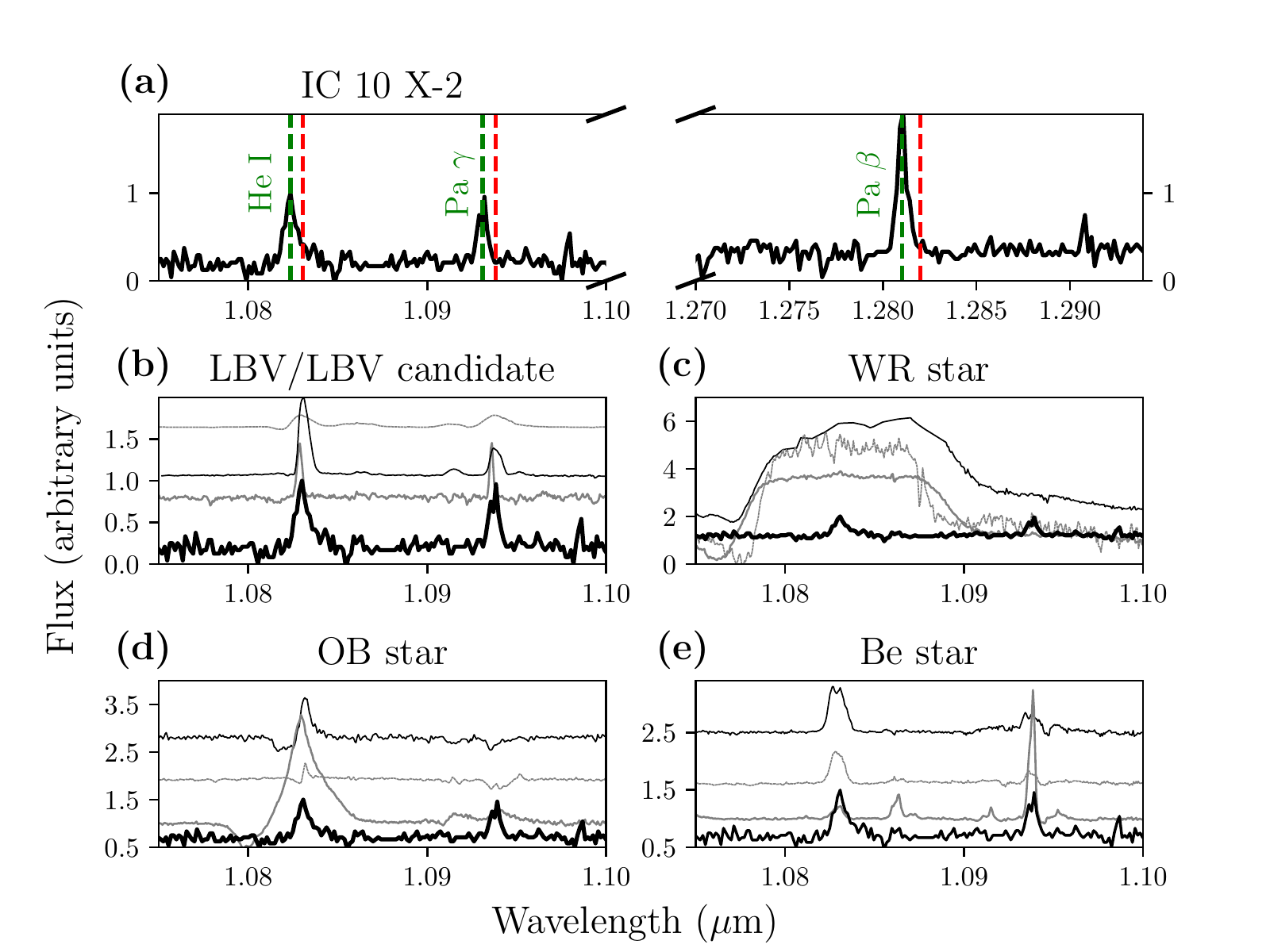}
        \caption{\textbf{(a)} He I, Paschen-Gamma (Pa $\gamma$), and Paschen-Beta (Pa $\beta$) emission lines detected with strong ($>$5) signal-to-noise ratio in the TripleSpec observations. The observed blueshifted and rest frame wavelengths for the three lines are indicated by the green and red dashed lines respectively. The latter two lines are close to the instrument resolution limit. The apparent double peak in the plot for Pa $\gamma$ is most likely an artifact. \textbf{(b)-(e)}  IC 10 X-2's rest-frame spectrum plotted with representative spectra of various stars from \cite{Groh2007}. The other lines, denoted with solid black, dashed grey, and solid grey, correspond to the following stars. \textbf{(b)} LBVs: W243 A2 Ia, HD 316286, and AG Carinae. \textbf{(c)} Wolf-Rayet stars: WR 90  WC7, WR 136  WN6(h), WR6 WN4. \textbf{(d)} OB stars: HD 152408  O8 Iafpe, HD 80077 B2 Ia+, HD 169454 B1 Ia+. \textbf{(e)} Be stars: $\chi$ Oph B1.5Ve, $\delta$ Sco B0.31Ve, $\chi$ Per B0Ve.}
        \label{fig:TripleSpecFeatures}
\end{figure*}

An IR light curve of IC 10 X-2 is shown in Fig. \ref{fig:IR_and_optical_lc}. Between the initial mid-IR observations taken 2004 July 23 and more recent observations taken from 2010 - 2016, the [3.6] and [4.5] flux densities from X-2 decrease factors of 0.3 ($\sim 2\sigma$) and 0.35 ($\sim 4\sigma$), respectively. Four months prior to the 2010 May X-ray outburst, $\lesssim$month-timescale variability is detected between 2010 Jan 29 and 2010 Feb 19 where the [3.6] and [4.5] fluxes decreased by factors of 0.30 and 0.35, respectively. 

Serendipitously, the three mid-IR observations preceding the outburst by $\sim2-4$ months were taken within $\sim$weeks - months of \textit{Chandra}/ACIS-S non-detections on 2010 Feb 11 and Apr 4 \citep{Laycock}. After the X-ray outburst, there is no obvious evidence of variability in the mid-IR with observations over the following $\sim6$ yr. Post-outburst, the [3.6] and [4.5] fluxes exhibit approximately constant values of 0.20 and 0.15 Jy, respectively.
%We consider \textbf{the mid-IR activity to } to be the IR counterpart of IC 10 X-2.
 
%The IR emission subsequently returns to quiescent levels after 2011. This gives us a timescale of about a month during which the neutron star absorbs the large amount of circumstellar wind dust required to power its fast, high-energy X-ray outburst.

%\textbf{[Couple paragraphs on the Optical light curve. Info to provide: median magnitudes, does the baseline flux stay constant over time? Compare to IR light curve.  want to find the average time between flares as well as the longest measured duration of a flare, also want to provide the amplitudes of the flares. Comment also on lack of flares in the most recent series of R-band observations--this is pretty weird.]}

Both the optical R- and g-band light curves of X-2 exhibit high amplitude, short-duration flares that deviate significantly ($>5\sigma$) from the quiescent brightness levels (Fig.~\ref{fig:IR_and_optical_lc}). The median, ``quiescent'' magnitudes in the R and g bands were $18.64 \pm 0.05$ and $20.17 \pm 0.08$ Vega mag, respectively. A Lomb-Scargle periodogram analysis of the R-band light curve provided no significant periodicity in the variability.

A flare was defined to be a brightening event with an amplitude greater than 5$\sigma$ above the quiescent magnitude. The brightest observed flare in the R-band occurred on 2013 Oct 10 (MJD 56575) and exhibited an amplitude of 1.32 mag. There were two detections of the 2013 Oct 10 flare taken 0.15 d apart, which provides a lower limit on the flare duration. Notably, this is the longest observed flare duration measured in the optical light curves. Upper limits were placed on the duration of each flare by defining the start and end of each flare to be magnitudes within or below 1$\sigma$ of the quiescent brightness. The average upper limit determined for each flare is 8.0 d.

%An upper limit on flare duration is determined by the time between the first observations preceding and following the flare that are within $1\sigma$ of the quiescent brightness or below it. Observations of the flare on 2013 Sept 15 (MJD 56550.4)

An average timescale between flare events of $15.1\pm 7.8$ d is derived for the R-band light curve of X-2. The $>60$ d gaps in the light curve with no coverage of X-2 were omitted from this calculation. Interestingly, there were no flares observed at R- nor g-Band between 2015 Aug 5 (MJD 57239.4) and the end of the observations on 2017 Feb 14 (MJD 57798.1)

Short-duration and significant dips in brightness ($>5\sigma$) are also apparent in both R- and g-band light curve in light curves of X-2. Only the dip on 2014 Jul 24 (MJD 56862.3) was detected with multiple consecutive observations, which provides a lower limit on the dip duration of $0.10$ d. The highest amplitude dip was observed on 2015 Jan 25 (MJD 57047.2) and decreased to 0.62 mag below the quiescent brightness.

%The median magnitudes in the R and g bands were $18.64 \pm 0.05$ and $20.17 \pm 0.08$ respectively. \textbf{Between 2013 Jul 07 and 2014 Feb 02, there were sufficient observations in the $R$-band to determine average flare properties. A flare was defined to be magnitudes brighter than 3$\sigma$ of the median magnitude. Lower limits on the duration of the flares were calculated for instances where there were 2 or more detections of the flare. Upper limits were placed on the duration of each flare, by defining the start and end of each flare to be magnitudes within or below 1$\sigma$ of the median. The durations between the brightest points in each flare were also calculated. The amplitude of each flare was calculated to be the difference between the brightest point and the simple average of the start and end points.}

%\textbf{The average time between flares in 2013 Jul 07 to 2014 Feb 02 was 18.48 days (10 samples). The average lower limit on the flares in the same time period was 2.23 days (6 samples), and the average upper limit was 8.00 days (11 samples). The average amplitude of the flares was 0.3759 magnitudes.}

\subsection{Near-IR Spectroscopy}
%%Present the information you derived about the line widths and centers, indicating which lines they correspond to. Show that the blueshift is consistent with that of IC 10
The TripleSpec instrument detected near-IR continuum emission from X-2 and three emission lines with high signal-to-noise ratio (see Fig. \ref{fig:TripleSpecFeatures}). We identified them as Helium I (He I, 1.0819 $\mu m$), Paschen-Gamma (Pa $\gamma$, 1.0926 $\mu m$), and Paschen-Beta (Pa $\beta$, 1.2805 $\mu$m). Comparing to their rest frame wavelengths at 1.0830 $\mu m$, 1.0933 $\mu m$, and 1.2815 $\mu m$ gives blueshifts of $-314$ km s$^{-1}$, $-332$ km s$^{-1}$, and $-307$ km s$^{-1}$, which are consistent with the $-340$ km s$^{-1}$ for IC 10 \citep{Laycock}. 

The center wavelengths and full-widths at half maximum (FWHM) were determined from Gaussian fits in CurveFit\textit{Mathematica}\footnote{http://www.sophphx.caltech.edu/CurveFit/}, a \textit{Mathematica} package. 	The FWHM were corrected for instrument resolution according to the equation:
\begin{equation}
	FWHM_{corrected} = \sqrt{FWHM_{obs}^2 - FWHM_{instr}^2}
\end{equation}
using the average spectral resolution $2600$, from the $2500-2700$ spectral resolution provided in TripleSpec instrument documentation.

The Pa $\gamma$ and Pa $\beta$ emission lines exhibit velocity widths near the 120 km s$^{-1}$ spectral resolution of TripleSpec. The velocity width of the He I 10830 \AA\ emission line, which provides a valuable diagnostic on the physical properties of stellar winds from massive stars (e.g. \citealt{Groh2007} is $\sim230$  km s$^{-1}$. This velocity is consistent with winds from quiescent LBVs \citep{Smith,humphreys2017} and the dense equatorial outflows from sgB[e] stars \citep{deWit} but slower than the $\gtrsim1000$ km s$^{-1}$ outflows exhibited by OB supergiants and WR stars.

\section{Discussion}

\subsection{On the Nature of the IR Counterpart: an LBV Candidate}
%\textbf{[Include paragraph on spectral lines identifying LBV/sgB[e] and ruling out other massive stars. Use line widths/velocities to further justify the LBV/sgB[e] claim]}

%The near-IR emission lines around He I 10830\AA\ provide a diagnostic on the nature and physical properties of X-2's IR counterpart. 

In order to discern the nature of X-2's stellar counterpart, we compare its near-IR spectum with the \citet{Groh2007} spectral atlas around He I 10830 \AA\ encompassing four categories of stars: (1) Be type, (2) LBV/LBV candidates (LBVc)/ex- or dormant LBVs, (3) OB supergiants, and (4) Wolf-Rayets. The $\sim230$ km s$^{-1}$ width of X-2's He I peak rules out the Wolf-Rayet class, which exhibits a broad ($\gtrsim1000$ km s$^{-1}$) He I emission line. Additionally, Wolf-Rayet stars are typically Hydrogen-poor which contradicts the clear presence of Hydrogen in X-2. X-2 has equally strong He I and Pa $\gamma$ lines, ruling out OB supergiants, in which the former is much stronger than the latter. In the remaining two categories, X-2 most resembles the LBVc W243 \citep{Clark} and the classical Be star HD 120991 that exhibit He I and Pa $\gamma$ lines of similar amplitude and width. It should be noted that LBVs exhibit spectroscopic variability, altering the relative strengths of the lines. 

\begin{figure*}[ht!]	\plotone{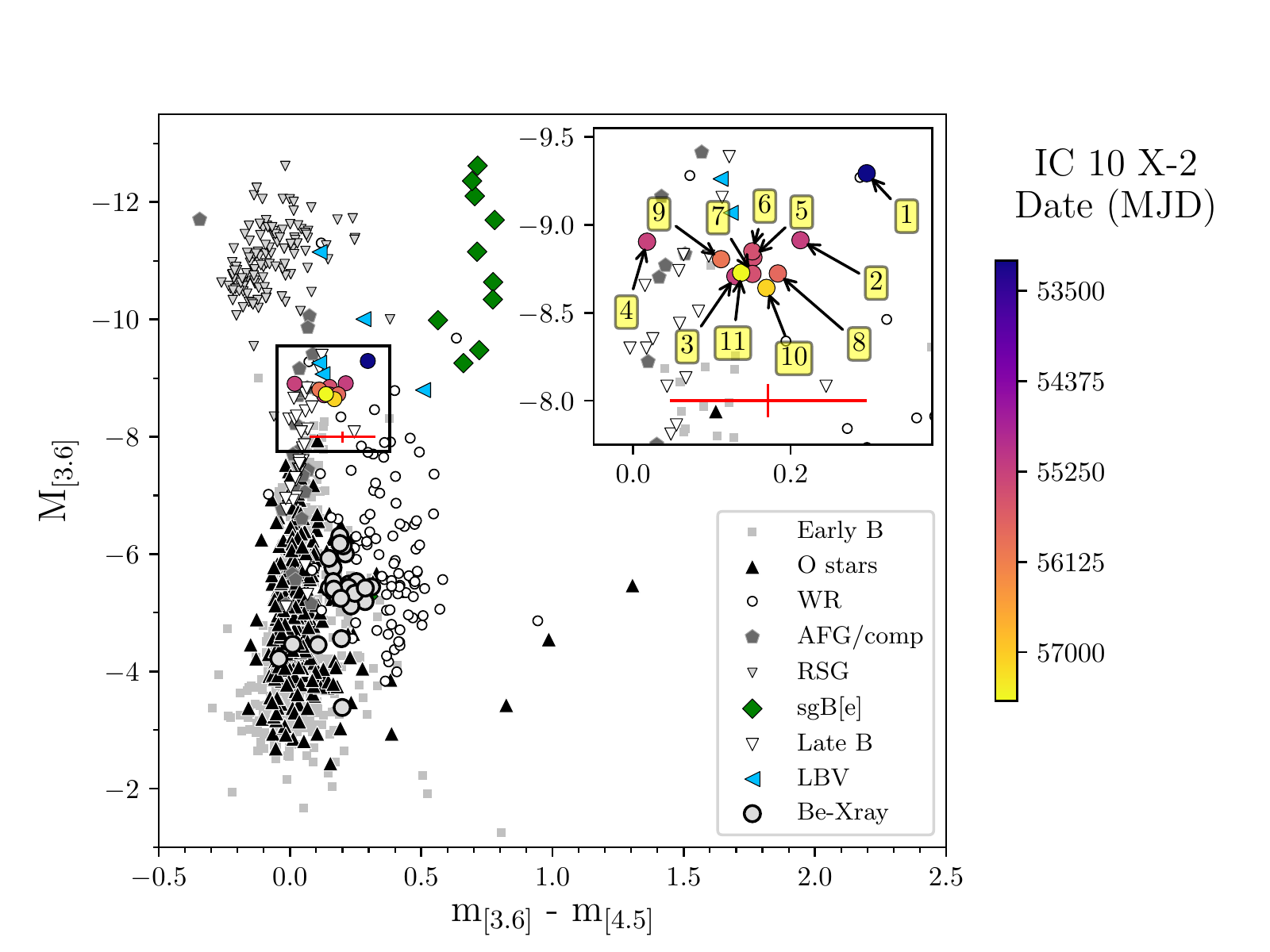}
	\caption{[3.6] - [4.5] color vs. [3.6] absolute magnitude of massive stars with IRAC counterparts in the Large Magellanic Cloud from Figure 2 of \citet{Bonanos}. IC 10 X-2 is plotted over eleven epochs spanning 2004 to 2016 (see color bar and inset). The error bar for the color and absolute magnitude of X-2 was calculated from uncertainties in the flux. The IC 10 distance modulus was taken to be 24.1 \citep{Laycock}.}
	\label{fig:BonanosCMD}
\end{figure*}

Another means of classifying the IR counterpart of IC 10 X-2 came from plotting its [3.6] color vs. [3.6] - [4.5] magnitude on a color-magnitude diagram (CMD) of known supergiant stars in the Large Magellanic Cloud (LMC) with \textit{Spitzer} counterparts shown in Fig. \ref{fig:BonanosCMD} of \citet{Bonanos}. The original color-magnitude diagram was recreated from online data published by \citet{Bonanos}. We used $\mu=18.91$ for the LMC distance modulus \citep{Bonanos}, and $\mu=24.1$ for the galaxy IC 10 \citep{Laycock}.
Throughout the observed mid-IR evolution, X-2 occupies a region in the mid-IR CMD consistent with A-, F-, G-type supergiants (AFG/comp in Fig.~\ref{fig:BonanosCMD}), late and early B-type supergiants, Wolf Rayets, and LBVs. Notably, X-2 is over two magnitudes brighter at 3.6 $\mu$m than the Be-XRB, which implies that the IR counterpart is unlikely a classical Be star. Wolf-Rayet and Early B-type supergiants are ruled out based on the properties of their near-IR emission lines discussed above. It is unlikely that X-2 is an AFG/late-B supergiant given the hardness of the radiation field required to produce the observed near-IR emission lines (Fig.~\ref{fig:TripleSpecFeatures}) noting also the absence of these feature in early B supergiants. We therefore interpret X-2 as an LBV/LBVc, which is supported by the presence of characteristic optical emission lines revealed by \citet{Laycock} (e.g. He II 4686).
%Although we cannot conclusively rule out an AFG/late-B supergiant interpretation for the IR counterpart, 

%\textbf{The presence of the He II 4686 emission line in optical spectra of X-2 \citep{Laycock} imply that the spectrum of the IR counterpart must be hard enough to ionize Helium, thus ruling out A-, F-, G-, and late B-type supergiants.}

LBVs typically exhibit luminosities greater than a few times $10^5$ L$_\odot$ \citep{Smith}. It is difficult to determine the bolometric luminosity of X-2 due to uncertainties in the line of sight extinction; however, \citet{Laycock} estimate extinction-corrected V-band magnitudes of X-2 as $M_V=-6.78$ for optically derived de-reddening and $M_V=-7.51$ for X-ray derived de-reddening. Assuming a bolometric correction of $-1.68$ mag, the approximate value for the LBV AG Carinae in 2002 March-July \citep{Groh2009}, we estimate the luminosity of X-2 to be $\sim2 - 4\times10^5$ L$_\odot$. This range of luminosities is consistent with the values exhibited by lower luminosity LBVs. We note that although the bolometric corrections will depend on the effective temperature of the star, there are LBVs that have absolute visual magnitudes consistent with X-2 ($M_V\sim-7$; \citealt{humphreys2014}).

%Applying a bolometric correction of $-0.2$ mag, the approximate value for the LBV AG Carinae \citep{Groh} \footnote{Note that the bolometric correction will vary depending on the effective temperature exhibited by the LBV}, 

%\textbf{[Add paragraph presenting the optical luminosity presented by Laycock et al. (2014) and perform the appropriate bolometric correction for LBVs to get a bolometric luminosity and compare that to other known LBVs. Comment on the result (most likely, it will be on the lower end of the LBV luminosity)]}

%The highly variable migration from redder to bluer colors suggests that the optical counterpart of IC 10 X-2 began as a sgB[e] star and is shifting to the LBV population. SgB[e] stars and LBVs exhibit similar properties but not the same spectroscopic and photometric variability, which could suggest that interactions with the compact object are affecting the evolution of the mass donor star.

\subsection{Mid-IR Free-Free Emission from Ionized Winds}

Massive stars with strong winds such as LBV and Wolf-Rayet stars can exhibit an IR excess due to free-free emission from their outflows (e.g. \citealt{Cohen1975}). However, an IR excess may also be attributed to the presence of warm circumstellar dust. We attempt to distinguish the nature of the mid-IR emission by performing a power-law fit, $F_\nu=C \,\lambda^{\alpha}$ mJy, to the 3.6, 4.5, 5.8 and 8.0 $\mu$m photometry taken 2004 Jul 23. The best-fit provides $C = 0.9\pm0.2$ mJy and $\alpha=-0.76\pm0.12$. This slope is consistent with the predicted $\alpha\sim-0.6$ value for free-free emission from an ionized outflow from a star with a spherical expansion and constant velocity (e.g. \citealt{Wright}).  The source of the mid-IR emission is unlikely due to dust given that color temperature estimates for the mid-IR emission at 3.6 and 4.5 $\mu$m indicate values of $\sim1700$ K, which is hotter than dust sublimation temperatures ($\sim1200$ K; \citealt{DraineBook}). We therefore interpret the IR excess from X-2 as free-free emission.

Work by \citet{humphreys2017} on distinguishing LBVs/LBVcs and supergiant B[e] stars shows that the IR spectral energy distribution from LBVs exhibit free-free emission with no evidence for warm circumstellar dust. Most supergiant B[e] stars however, are found to have warm circumstellar dust. \citet{humphreys2017}  also indicate that dusty/non-dusty LBV/sgB[e] stars are clearly separated in mid-IR color space, where stars with $[3.6]-[4.5]<0.5$ do not show warm circumstellar dust. This distinction between sgB[e] and LBV stars is also noticeable in Fig.~\ref{fig:BonanosCMD} and supports the interpretation of X-2 as an LBV/LBVc.  

%\textbf{[Include paragraph on what the emission is, why it's not likely dust (can reference Humphreys et al. 2016 since the mid-IR colors don't indicate dust). Mention that outflows from massive stars can produce free-free emission (e.g.Wright \& Barlow 1975, Fuchs et al. 2006, Mauerhan et al. 2011) and show that the power-law fit is consistent with the power-law predicted from free-free emission for winds from massive stars--can also cite Humphreys et al. 2016 mentioning that this is consistent with LBVs.]}

If the mid-IR flux originates from free-free emission in the winds of IC 10 X-2's optical counterpart, the fluxes can be used to estimate its mass-loss rate (e.g. \citealt{Fuchs}). Adopting the formalism of \citet{Wright} and assuming a thick geometry for the wind (i.e. the ratio of structural size scales in the outflow do not exceed $\sim10$), the mass-loss rate, $\dot{M}_w$, is related to the mid-IR flux at $4.5$ $\mu$m, $F_{4.5}$, and other outflow properties as follows:

\begin{eqnarray}
\dot{M}_w\sim6\times10^{-5}\frac{\mu}{\gamma_e^{1/2}\,g^{1/2}\,Z}\left(\frac{d}{660\,\mathrm{kpc}}\right)^{3/2} \cdot \nonumber \\
\left(\frac{F_{4.5}}{174\,\mu\mathrm{Jy}}\right)^{3/4}
\left(\frac{v_{w}}{200\,\mathrm{km}\,\mathrm{s}^{-1}}\right)\,\,\mathrm{M}\odot\,\mathrm{yr}^{-1},
\label{eq:FF}
\end{eqnarray}
\noindent
where $\mu=1$ is the mean molecular weight per nucleon, $d$ is the distance to IC 10 X-2, v$_w$ is the wind velocity, $\gamma_e=1$ is the number of free electrons per nucleon, $g=1$ is the Gaunt factor, $Z=1$ is the mean charge per ion. The mass-loss rates determined by Eq.~\ref{eq:FF} from the range of fluxes exhibited by IC 10 X-2 at 4.5 $\mu$m is $\dot{M}_w\approx3\times10^{-5}-10^{-4}$ M$_\odot$ yr$^{-1}$ , which is consistent with the mass-loss rates exhibited by LBVs in quiescence (e.g. \citealt{Smith}). 

%\textbf{It is interesting to draw comparisons to the mid-IR behavior of the X-ray outburst from the sgB[e]-HMXB SN~2010da, where the 3.6 and 4.5 $\mu$m emission exhibited an exponential increase 5 months prior to the outburst \citep{Lau}. However, the mid-IR emission from SN~2010da was attributed dust produced by the sgB[e] companion, whereas the mid-IR emission from X-2 likely originates from its ionized winds. Due to the short $\lesssim$month-timescale variability exhibit by both the mid-IR and X-ray emission and the lack of contemporaneous multi-wavelength coverage, it is difficult to determine whether or not the mid-IR and X-ray variability originate from the same mechanism(s). Further multi-wavelengths observations are required to explore the possible connection between the mid-IR and X-ray activity from X-2.}

\subsection{Interpreting the Optical and X-Ray Outbursts}

Since the optical and X-ray data were not acquired contemporaneously, it is difficult to determine whether or not the X-ray and optical variability share the same origin. However, the $\sim 0.15$ d ($\sim 4$ hr) lower-limit on the optical variability timescale is consistent with observed durations of X-ray flares from known SFXTs (a few hours; \citealt{Romano}).

We can address whether the short timescale ($0.15$ d $\lesssim \Delta t \lesssim 4$ d) and high amplitude ($\sim 1$ mag) variability exhibited by X-2 (Fig.~\ref{fig:IR_and_optical_lc}) is due to activity from the stellar counterpart or linked to the binary nature of the system. LBVs and OB supergiants are known to exhibit optical variability on similarly short timescales of days to weeks, but the brightness only varies by up to a few tenths of a magnitude (e.g. Bresolin et al. 2004). Alternatively, LBVs are known to undergo phases of extreme photometric variability from 1 - 2 mag but on timescales of years to decades. The short timescale, high amplitude variability from X-2 is therefore most likely due to the presence of the compact object and linked to the X-ray variability.

There are several mechanisms considered for SFXTs that could account for the short optical flares from X-2. The flares may be due to accretion of clumpy winds from the donor star, motion of the compact object through an asymmetric outflow, and/or ``gated'' accretion due to magnetic processes (see \citealt{Sidoli2013} and ref. therein). With the available data on X-2, it is difficult to address the later mechanism; however, we can study the first two in the theoretical framework of Bondi-Hoyle accretion (e.g. \citealt{Davidson}). 

Assuming an accreting compact object of mass M$_{CO}$ is traveling at a velocity v$_{rel}$ relative to the winds from the supergiant donor star of mass M$_*$ with a mass-loss rate $\dot{M}_w$ and outflow velocity v$_w$, the accretion rate onto the compact object can be approximated by 

\begin{equation}
\dot{M}_\mathrm{acc}=\zeta\frac{(G\,M_\mathrm{CO})^2}{v_\mathrm{rel}^3}\frac{\dot{M}_\mathrm{w}}{a^2\,v_\mathrm{w}},
\label{eq:Lx2}
\end{equation}

\noindent
where $a$ is the orbital separation between the compact object and donor star, and $\zeta$ is the correction factor for radiation pressure and the finite gas cooling time (assumed to be $\zeta=1$, e.g. \citealt{Oskinova}). The relative velocity between the winds and the orbital motion of the compact object moving at a speed of v$_{CO}$ is 

\begin{equation}
v_{rel}=\sqrt{v_{CO}^2+v_{w}^2},
\end{equation}
\noindent
where v$_{CO}$ can be approximated by v$_{CO}\approx \sqrt{G M_*/a}$ assuming nearly circular orbits and $M_*>>M_\mathrm{CO}$. The X-ray luminosity from the accreting compact object is then

\begin{equation}
L_X=\eta\, \dot{M}_{acc}c^2\approx\eta\,\zeta\frac{(G\,M_\mathrm{CO})^2}{(\frac{G M_*}{a}+v_w^2)^{3/2}}\frac{\dot{M}_\mathrm{w}}{a^2\,v_\mathrm{w}}c^2,
\label{eq:BHA}
\end{equation}

\noindent
where $\eta$ is an efficiency factor that depends on the accretion physics and is assumed to be $\eta\sim 0.1$ (e.g. \citealt{Oskinova}). 

In order to test the viability of Eq.~\ref{eq:BHA}, we use it to compare against the orbital period ($P_\mathrm{orb}=8.95$ d) of the well-studied wind-fed HMXB Vela X-1 (\citealt{Hiltner1972}; \citealt{Lamers1976}; \citealt{Quaintrell2003}). We adopt the Vela X-1 supergiant stellar wind velocity, mass, and mass-loss rate provided by \citet{GG2016} and the X-ray luminosity derived by \citet{Sako1999}: $v_w=264$ km s$^{-1}$ $M_*=21.5$ M$_\odot$, $\dot{M}=6.3\times10^{-7}$ M$_\odot$ yr$^{-1}$, and L$_X=4.5\times10^{36}$ erg s$^{-1}$. Assuming the mass of the accreting neutron star is $M_\mathrm{CO}=1.5$ M$_\odot$, we estimate from Eq.~\ref{eq:BHA} and Kepler's 3rd Law that $P_\mathrm{orb}\sim 16$ d, which is within a factor of 2 of the observed orbital period for Vela X-1. Equation~\ref{eq:BHA} should therefore provide reasonable estimates or limits for the orbital parameters of wind-fed X-ray binaries in nearly circular orbits.

First, we consider the scenario where the optical/X-ray flares were due to the compact object encountering an asymmetric outflow from the stellar counterpart. Such mass-loss asymmetries are naturally inferred for stars that are observed to be rapidly rotating like bona-fide galactic LBVs (e.g. \citealt{Groh2009b}). Adopting the derived mass-loss rate of $\dot{M}_w\sim5\times10^{-5}$ M$_\odot$ yr$^{-1}$ and wind velocity v$_w\sim200$ km s$^{-1}$, and assuming a donor star mass of $M_*\sim30$ M$_\odot$ and compact object mass of M$_{CO}\sim1.5$ M$_\odot$, we find from Eq.~\ref{eq:BHA} and Kepler's 3rd Law that for the range of X-ray luminosities exhibited by X-2 in outburst, $L_X= 3.4\times10^{36}-1.8\times10^{37}$ erg s$^{-1}$,

\begin{equation}
\,P_{orb}\sim 500-8600\,\,\mathrm{d}.
\end{equation}

\noindent
The derived orbital timescales under the asymmetric outflow scenario are over an order of magnitude longer than the observed timescales between optical flares ($\sim15$ d; Sec.~\ref{subsec:lc}). The timing of the flares in this scenario should also be periodic since they correspond to the orbital motion of the compact object through the stellar outflow. The lack of any significant periodicities identified in a Lomb-Scargle periodogram analysis of the X-2 optical light and the inconsistency between the theoretical and observed flare timescales suggests that the SFXT flares from X-2 did not originate from interactions between the compact object and an asymmetric outflow. 

%Assuming that the optical flares correspond to the observed X-ray transient events, it is therefore unlikely that the flares originate from interaction of the compact object with an asymmetric outflow due to the inconsistent timescales. The lack of any significant periodicities in a Lomb-Scargle periodogram analysis of the X-2 optical light curve supports this conclusion.}

We now consider the scenario were the optical/X-ray flares from X-2 were due to accretion of clumps in an inhomogeneous outflow from the stellar counterpart. Clumping in the winds of massive stars has been observationally confirmed and is also backed by stellar wind theory (e.g. \citealt{Oskinova}). With the observational constraints on the flare duration, $\Delta t\sim 0.2 - 8$ d, and the time between flares, $T\sim 15$ d, the clump volume filling factor, $f_V$, can be estimated for the stellar outflow of X-2 and compared with the $f_V$ of supergiant winds.

Assuming a simple model where the outflows consist entirely of clumps with radius $R_{cl}$ and mass $M_{cl}$, the clump volume filling factor is given by the ratio of the homogeneous wind density, $\rho_h$, and the individual clump density, $\rho_{cl}$: $f_V=\frac{\rho_h}{\rho_{cl}}$. The homogeneous wind and clump densities can be written as

\begin{equation}
\rho_h=\frac{\dot{M_{cl}}}{4\pi\,R^2\,v_{cl}}\,\,\mathrm{and}\,\,\rho_{cl}=\frac{M_{cl}}{4/3\,\pi\,R_{cl}^3},
\label{eq:densities}
\end{equation}

\noindent
respectively, where $\dot{M_{cl}}$ is mass-loss from the clumped outflow, $R$ is the distance between the stellar counterpart and compact object, and $v_{cl}$ is the clump outflow velocity. $\dot{M_{cl}}$ can be related to the timescale between the flares, $T$, as follows (see Walter \& Zurita Heras 2007):

\begin{equation}
\dot{M_{cl}}=\frac{4\,R^2 M_{cl}}{R_{cl}^2\,T}.
\label{eq:mdot}
\end{equation}

\noindent
By combining Eq.~\ref{eq:densities} and~\ref{eq:mdot}, $f_V$ can then be expressed as

\begin{equation}
f_V=\frac{4\,R_{cl}}{3\,T\,v_{cl}}\sim \frac{4\,\Delta t}{3\,T},
\label{eq:fV}
\end{equation}

\noindent
where we assumed $R_{cl}\sim v_{cl}\, \Delta t$. Given the observed constraints for X-2 on $\Delta t\sim 0.2 - 8$ d and $T\sim 15$ d, we determine from Eq.~\ref{eq:fV} that 

\begin{equation}
0.01<f_V< 0.71.
\end{equation}

\noindent
Although the upper limit of $f_v$ is not well constrained from our observations, the range of values for $f_V$ overlaps with observationally derived clump volume filling factors for massive star winds: $f_V\sim0.02-0.1$ (e.g. \citealt{Bouret}; \citealt{Oskinova2007}). The additional evidence of ``eclipses" in the light curve, which may arise from optically thick clumps in the winds of the optical counterpart (Fig.~\ref{fig:IR_and_optical_lc}), supports the interpretation of the optical flares as clump accretion onto the compact object.

\section{Conclusion}
We report time-varying characteristics of the optical and IR counterpart to IC 10 X-2, a supergiant fast X-ray transient candidate discovered by its X-ray outbursts in 2004 and 2010~\citep{Laycock}. We utilize observations taken by \textit{Spitzer} in the mid-IR, WIRC and TripleSpec in the near-IR, and iPTF at optical bands. Although there are no concurrent observations with the X-ray transient events from X-2, the optical and IR data provide valuable information on the nature of the stellar counterpart and its outflow that may be contributing to the X-ray flares. 

Near-IR spectroscopy of X-2 reveal moderately resolved ($\sim200$ km s$^{-1}$) hydrogen and helium emission lines that rule out stellar counterparts that exhibit high wind velocities like OB supergiants and WR stars. The presence of these lines also rules out the stars that exhibit low ionizing fluxes like A, F, G, or late B supergiants. The spectra more closely resemble the properties of an LBV/LBVc. This LBV/LBVc interpretation is supported by the similar mid-IR color and magnitude X-2 shares with LBVs observed in the LMC (Fig.~\ref{fig:IC10X2Color}) and the $6\times10^{-5}$ M$_\odot$ yr$^{-1}$ mass-loss rate estimated from the mid-IR flux assuming it originates from free-free emission in ionized winds. This mass-loss rate is consistent with LBVs in quiescence. 

The second key result from our analysis is the presence of short ($\sim$ day) high amplitude ($\sim 1$ mag) optical flares and eclipses in the iPTF light curves taken between 2009 and 2017. We interpret the origin of the flares as accretion of optically thick clumps formed in the winds of the stellar companion in X-2 on to the compact object. Constraints on the clump volume filling factor in the stellar winds as determined by the timing and duration of the optical flares show a range of values that overlap with the observed clump volume filling factor in massive star winds. 

Currently, only 12 SFXTs \citep{Romano} and $\sim20$ bona-fide LBVs are known. X-2, as an SFXT candidate hosting an LBV/LBVc, is therefore at the cross section of these two rare classes of sources. Future coordinated, multi-wavelength observations of X-2 present a unique opportunity to study the accretion processes in SFTXs and the interactions between the compact object and the mass-loss from an LBV/LBVc companion.
\acknowledgments
This project received funding from the California Institute of Technology Student-Faculty Program's Summer Undergraduate Research Fellowship (SURF) and the Flintridge Foundation. S. Kwan would like to thank M. Kasliwal, R. Lau, and J. Jencson for their mentorship. R. Lau thanks M. Heida for the valuable discussion on SFXTs. We thank the \textit{Spitzer} InfraRed Intensive Transient Survey team and Palomar Observatory for their scientific support. Palomar Observatory is owned and operated by the California Institute of Technology, and administered by Caltech Optical Observatories.

This publication makes use of data products from the Two Micron All Sky Survey, which is a joint project of the University of Massachusetts and the Infrared Processing and Analysis Center/California Institute of Technology, funded by the National Aeronautics and Space Administration and the National Science Foundation.

This research has made use of the SVO Filter Profile Service (http://svo2.cab.inta-csic.es/theory/fps/) supported from the Spanish MINECO through grant AyA2014-55216.

%% for the paper.  Each keyword is check against the master list during
%% copy editing.  Individual instruments can be provided in parentheses,
%% after the keyword, but they are not verified.

\vspace{5mm}
\facilities{P200(WIRC, TripleSpec), Spitzer IRAC}

\software{SAO DS9, Python, Mathematica}
\pagebreak

\begin{deluxetable}{cc}
\tablecaption{R-band Photometry}
\tablewidth{0pt}
\tablehead{Date(MJD) & Mag}
\startdata
55514.23616	&	$	17.86	\pm	0.02	$	\\
56488.32939	&	$	18.76	\pm	0.03	$	\\
56488.36582	&	$	18.76	\pm	0.03	$	\\
56489.32626	&	$	18.78	\pm	0.04	$	\\
56489.36289	&	$	18.70	\pm	0.03	$	\\
56491.35721	&	$	18.65	\pm	0.04	$	\\
56498.2898	&	$	18.40	\pm	0.03	$	\\
56500.43422	&	$	18.53	\pm	0.01	$	\\
56501.41988	&	$	18.77	\pm	0.03	$	\\
56501.45605	&	$	18.73	\pm	0.03	$	\\
56502.46742	&	$	18.56	\pm	0.04	$	\\
56503.27465	&	$	18.66	\pm	0.03	$	\\
56503.31132	&	$	18.61	\pm	0.03	$	\\
56505.27163	&	$	18.49	\pm	0.02	$	\\
56505.41425	&	$	18.58	\pm	0.03	$	\\
56513.5047	&	$	18.78	\pm	0.06	$	\\
56515.4967	&	$	18.65	\pm	0.03	$	\\
56516.20826	&	$	18.66	\pm	0.03	$	\\
56516.24802	&	$	18.66	\pm	0.02	$	\\
56518.20242	&	$	18.49	\pm	0.04	$	\\
56518.25131	&	$	18.62	\pm	0.03	$	\\
56518.29041	&	$	18.64	\pm	0.03	$	\\
56520.19334	&	$	18.65	\pm	0.04	$	\\
56520.25701	&	$	18.61	\pm	0.04	$	\\
56520.29488	&	$	18.68	\pm	0.04	$	\\
56529.39718	&	$	18.65	\pm	0.03	$	\\
56529.43346	&	$	18.72	\pm	0.04	$	\\
56529.4695	&	$	18.77	\pm	0.06	$	\\
56532.1597	&	$	18.51	\pm	0.04	$	\\
56532.24578	&	$	18.51	\pm	0.02	$	\\
56532.31912	&	$	18.48	\pm	0.02	$	\\
56537.16429	&	$	18.68	\pm	0.03	$	\\
56537.45612	&	$	18.73	\pm	0.03	$	\\
56537.4925	&	$	18.67	\pm	0.02	$	\\
56539.39483	&	$	18.71	\pm	0.02	$	\\
56539.43118	&	$	18.67	\pm	0.02	$	\\
56539.46814	&	$	18.68	\pm	0.02	$	\\
56541.40691	&	$	18.46	\pm	0.03	$	\\
56541.44414	&	$	18.57	\pm	0.02	$	\\
56543.34669	&	$	18.65	\pm	0.02	$	\\
56543.43679	&	$	18.68	\pm	0.02	$	\\
56545.33896	&	$	18.80	\pm	0.04	$	\\
56546.14125	&	$	18.75	\pm	0.04	$	\\
56546.24068	&	$	18.75	\pm	0.03	$	\\
56546.41507	&	$	18.83	\pm	0.02	$	\\
56548.318	&	$	18.64	\pm	0.03	$	\\
56548.45631	&	$	18.58	\pm	0.02	$	\\
56548.49296	&	$	18.56	\pm	0.03	$	\\
56550.40024	&	$	18.04	\pm	0.02	$	\\
56550.44832	&	$	17.99	\pm	0.02	$	\\
56550.48191	&	$	18.13	\pm	0.03	$	\\
56552.38512	&	$	18.51	\pm	0.04	$	\\
56552.4323	&	$	18.68	\pm	0.04	$	\\
56552.46575	&	$	18.55	\pm	0.04	$	\\
56558.12679	&	$	18.82	\pm	0.04	$	\\
56558.1594	&	$	18.77	\pm	0.04	$	\\
56558.19295	&	$	18.75	\pm	0.04	$	\\
56560.11681	&	$	18.68	\pm	0.04	$	\\
56560.15059	&	$	18.72	\pm	0.03	$	\\
56560.20559	&	$	18.67	\pm	0.04	$	\\
56562.40247	&	$	18.88	\pm	0.04	$	\\
56562.47419	&	$	18.80	\pm	0.05	$	\\
56564.38602	&	$	18.75	\pm	0.04	$	\\
56564.46833	&	$	18.70	\pm	0.03	$	\\
56566.37091	&	$	18.63	\pm	0.04	$	\\
56566.44337	&	$	18.73	\pm	0.04	$	\\
56568.34645	&	$	18.73	\pm	0.04	$	\\
56568.41987	&	$	18.76	\pm	0.03	$	\\
56570.32339	&	$	18.87	\pm	0.07	$	\\
56570.46095	&	$	18.71	\pm	0.12	$	\\
56572.36285	&	$	18.83	\pm	0.04	$	\\
56572.44725	&	$	18.81	\pm	0.05	$	\\
56576.33208	&	$	17.48	\pm	0.02	$	\\
56576.48402	&	$	17.32	\pm	0.02	$	\\
56580.32748	&	$	18.64	\pm	0.05	$	\\
56580.35282	&	$	18.64	\pm	0.03	$	\\
56587.12142	&	$	18.21	\pm	0.02	$	\\
56587.25208	&	$	18.23	\pm	0.02	$	\\
56589.15391	&	$	18.51	\pm	0.03	$	\\
56589.18001	&	$	18.46	\pm	0.02	$	\\
56591.09043	&	$	18.56	\pm	0.03	$	\\
56591.11758	&	$	18.54	\pm	0.03	$	\\
56593.10497	&	$	18.68	\pm	0.04	$	\\
56593.13542	&	$	18.72	\pm	0.04	$	\\
56596.09009	&	$	18.73	\pm	0.03	$	\\
56596.29199	&	$	18.74	\pm	0.05	$	\\
56598.19435	&	$	18.72	\pm	0.03	$	\\
56598.27627	&	$	18.69	\pm	0.03	$	\\
56600.19255	&	$	18.81	\pm	0.04	$	\\
56600.35051	&	$	18.79	\pm	0.05	$	\\
56602.25343	&	$	18.36	\pm	0.07	$	\\
56602.27883	&	$	18.34	\pm	0.05	$	\\
56604.23115	&	$	18.48	\pm	0.03	$	\\
56604.40069	&	$	18.37	\pm	0.02	$	\\
56607.08198	&	$	18.65	\pm	0.04	$	\\
56607.28429	&	$	18.63	\pm	0.03	$	\\
56609.17745	&	$	18.66	\pm	0.04	$	\\
56611.10898	&	$	18.64	\pm	0.04	$	\\
56611.13329	&	$	18.64	\pm	0.04	$	\\
56621.29479	&	$	18.61	\pm	0.04	$	\\
56622.08446	&	$	18.66	\pm	0.03	$	\\
56622.10905	&	$	18.68	\pm	0.04	$	\\
56625.20354	&	$	18.76	\pm	0.03	$	\\
56625.23247	&	$	18.74	\pm	0.04	$	\\
56627.30289	&	$	18.72	\pm	0.05	$	\\
56637.08132	&	$	18.58	\pm	0.05	$	\\
56637.10642	&	$	18.53	\pm	0.03	$	\\
56639.10692	&	$	18.55	\pm	0.04	$	\\
56648.0828	&	$	18.54	\pm	0.03	$	\\
56648.10917	&	$	18.56	\pm	0.04	$	\\
56653.09568	&	$	18.75	\pm	0.07	$	\\
56653.13734	&	$	18.64	\pm	0.05	$	\\
56659.08707	&	$	18.42	\pm	0.03	$	\\
56659.1119	&	$	18.41	\pm	0.03	$	\\
56663.09284	&	$	18.73	\pm	0.03	$	\\
56663.11843	&	$	18.65	\pm	0.04	$	\\
56666.08588	&	$	18.43	\pm	0.04	$	\\
56668.0877	&	$	18.50	\pm	0.06	$	\\
56668.11222	&	$	18.38	\pm	0.04	$	\\
56670.08942	&	$	18.56	\pm	0.08	$	\\
56675.09919	&	$	18.74	\pm	0.04	$	\\
56675.14168	&	$	18.63	\pm	0.03	$	\\
56677.21062	&	$	18.67	\pm	0.05	$	\\
56680.10995	&	$	18.80	\pm	0.04	$	\\
56680.1374	&	$	18.78	\pm	0.05	$	\\
56683.09992	&	$	17.84	\pm	0.02	$	\\
56683.13194	&	$	17.86	\pm	0.01	$	\\
56685.1092	&	$	18.36	\pm	0.02	$	\\
56687.10219	&	$	18.44	\pm	0.03	$	\\
56687.14475	&	$	18.52	\pm	0.02	$	\\
56690.1052	&	$	18.38	\pm	0.03	$	\\
56690.15279	&	$	18.42	\pm	0.04	$	\\
56697.11039	&	$	18.72	\pm	0.03	$	\\
56771.50559	&	$	18.80	\pm	0.11	$	\\
56772.50256	&	$	18.51	\pm	0.08	$	\\
56775.49464	&	$	18.27	\pm	0.03	$	\\
56776.49197	&	$	18.41	\pm	0.09	$	\\
56777.48838	&	$	18.58	\pm	0.12	$	\\
56785.46738	&	$	18.61	\pm	0.04	$	\\
56785.49089	&	$	18.54	\pm	0.06	$	\\
56789.45628	&	$	19.01	\pm	0.12	$	\\
56789.48222	&	$	18.55	\pm	0.06	$	\\
56837.47402	&	$	18.68	\pm	0.02	$	\\
56845.45242	&	$	18.03	\pm	0.04	$	\\
56845.47884	&	$	18.32	\pm	0.02	$	\\
56855.46222	&	$	18.63	\pm	0.04	$	\\
56862.27785	&	$	19.07	\pm	0.15	$	\\
56862.3165	&	$	19.07	\pm	0.22	$	\\
56862.3553	&	$	18.42	\pm	0.07	$	\\
56862.38817	&	$	18.34	\pm	0.07	$	\\
56869.37086	&	$	18.42	\pm	0.02	$	\\
56869.40135	&	$	18.08	\pm	0.04	$	\\
56869.43428	&	$	18.33	\pm	0.02	$	\\
56869.46248	&	$	17.77	\pm	0.06	$	\\
56869.49927	&	$	18.31	\pm	0.03	$	\\
56876.41522	&	$	18.67	\pm	0.02	$	\\
56876.44466	&	$	18.63	\pm	0.02	$	\\
56876.45058	&	$	18.51	\pm	0.03	$	\\
56876.48899	&	$	18.37	\pm	0.03	$	\\
56882.38472	&	$	18.49	\pm	0.02	$	\\
56882.45426	&	$	18.54	\pm	0.03	$	\\
56886.40657	&	$	18.81	\pm	0.03	$	\\
56886.43473	&	$	18.66	\pm	0.04	$	\\
56886.46407	&	$	18.77	\pm	0.03	$	\\
56886.48626	&	$	18.62	\pm	0.04	$	\\
56886.49722	&	$	18.68	\pm	0.03	$	\\
56886.50936	&	$	18.71	\pm	0.04	$	\\
56932.49492	&	$	18.53	\pm	0.04	$	\\
56932.52231	&	$	18.61	\pm	0.10	$	\\
56940.10688	&	$	18.66	\pm	0.07	$	\\
56972.2089	&	$	18.73	\pm	0.04	$	\\
56972.23176	&	$	18.78	\pm	0.05	$	\\
56979.21826	&	$	18.38	\pm	0.05	$	\\
56980.21221	&	$	18.47	\pm	0.03	$	\\
56986.22259	&	$	18.60	\pm	0.04	$	\\
56986.28578	&	$	18.53	\pm	0.05	$	\\
56987.20258	&	$	18.78	\pm	0.03	$	\\
56987.26433	&	$	18.51	\pm	0.13	$	\\
56988.16044	&	$	18.30	\pm	0.05	$	\\
56989.10013	&	$	18.58	\pm	0.04	$	\\
56989.16557	&	$	18.61	\pm	0.03	$	\\
56990.09579	&	$	18.50	\pm	0.08	$	\\
56990.15965	&	$	18.52	\pm	0.02	$	\\
57000.12383	&	$	18.48	\pm	0.05	$	\\
57000.15459	&	$	18.53	\pm	0.04	$	\\
57001.10524	&	$	18.44	\pm	0.03	$	\\
57001.13257	&	$	18.50	\pm	0.02	$	\\
57011.12233	&	$	18.47	\pm	0.03	$	\\
57011.14887	&	$	18.55	\pm	0.03	$	\\
57014.10122	&	$	18.73	\pm	0.04	$	\\
57014.12328	&	$	18.56	\pm	0.05	$	\\
57015.09928	&	$	18.81	\pm	0.05	$	\\
57015.11895	&	$	18.17	\pm	0.13	$	\\
57018.10068	&	$	18.54	\pm	0.04	$	\\
57018.12509	&	$	18.58	\pm	0.04	$	\\
57019.09567	&	$	19.15	\pm	0.15	$	\\
57019.10936	&	$	18.74	\pm	0.05	$	\\
57030.09887	&	$	18.52	\pm	0.04	$	\\
57030.16275	&	$	18.55	\pm	0.04	$	\\
57036.11696	&	$	18.87	\pm	0.08	$	\\
57037.10262	&	$	18.49	\pm	0.03	$	\\
57037.15286	&	$	18.53	\pm	0.04	$	\\
57038.15338	&	$	18.80	\pm	0.15	$	\\
57039.10544	&	$	18.59	\pm	0.03	$	\\
57039.13707	&	$	18.55	\pm	0.02	$	\\
57040.10478	&	$	18.49	\pm	0.03	$	\\
57040.15436	&	$	18.85	\pm	0.11	$	\\
57041.14765	&	$	18.38	\pm	0.03	$	\\
57042.10102	&	$	18.47	\pm	0.02	$	\\
57042.1477	&	$	17.63	\pm	0.05	$	\\
57043.10185	&	$	18.55	\pm	0.14	$	\\
57043.14678	&	$	18.54	\pm	0.03	$	\\
57046.10066	&	$	18.35	\pm	0.05	$	\\
57046.14241	&	$	18.82	\pm	0.09	$	\\
57047.15525	&	$	19.27	\pm	0.29	$	\\
57059.10799	&	$	18.83	\pm	0.04	$	\\
57059.11758	&	$	18.55	\pm	0.04	$	\\
57063.10801	&	$	18.56	\pm	0.04	$	\\
57063.12001	&	$	18.23	\pm	0.07	$	\\
57064.1092	&	$	18.46	\pm	0.07	$	\\
57064.12826	&	$	19.03	\pm	0.09	$	\\
57168.46631	&	$	18.65	\pm	0.07	$	\\
57179.4764	&	$	18.95	\pm	0.11	$	\\
57180.45855	&	$	18.79	\pm	0.04	$	\\
57185.46873	&	$	18.70	\pm	0.07	$	\\
57198.43492	&	$	18.85	\pm	0.08	$	\\
57198.46803	&	$	18.44	\pm	0.08	$	\\
57207.48208	&	$	18.62	\pm	0.03	$	\\
57210.43863	&	$	18.61	\pm	0.03	$	\\
57210.47753	&	$	18.46	\pm	0.07	$	\\
57214.46127	&	$	18.81	\pm	0.05	$	\\
57218.42646	&	$	18.76	\pm	0.03	$	\\
57218.47632	&	$	18.75	\pm	0.02	$	\\
57230.4439	&	$	18.88	\pm	0.05	$	\\
57237.43162	&	$	18.58	\pm	0.03	$	\\
57237.48108	&	$	18.26	\pm	0.07	$	\\
57239.42503	&	$	17.90	\pm	0.04	$	\\
57239.47012	&	$	18.48	\pm	0.05	$	\\
57242.35853	&	$	18.44	\pm	0.05	$	\\
57242.40331	&	$	18.14	\pm	0.04	$	\\
57245.30272	&	$	18.48	\pm	0.03	$	\\
57245.33991	&	$	18.28	\pm	0.05	$	\\
57248.3374	&	$	18.37	\pm	0.05	$	\\
57248.37045	&	$	18.48	\pm	0.02	$	\\
57251.32752	&	$	18.65	\pm	0.05	$	\\
57251.35716	&	$	18.74	\pm	0.04	$	\\
57267.43427	&	$	18.65	\pm	0.03	$	\\
57267.47725	&	$	18.64	\pm	0.04	$	\\
57363.09105	&	$	18.95	\pm	0.07	$	\\
57369.08424	&	$	18.67	\pm	0.04	$	\\
57369.0931	&	$	18.53	\pm	0.04	$	\\
57374.08289	&	$	18.53	\pm	0.03	$	\\
57374.09191	&	$	18.44	\pm	0.03	$	\\
57400.09411	&	$	18.60	\pm	0.04	$	\\
57400.10291	&	$	19.18	\pm	0.08	$	\\
57425.1106	&	$	18.56	\pm	0.13	$	\\
57431.10759	&	$	18.53	\pm	0.04	$	\\
57431.11384	&	$	18.56	\pm	0.04	$	\\
57434.12007	&	$	18.53	\pm	0.05	$	\\
57434.12381	&	$	18.58	\pm	0.03	$	\\
57544.48028	&	$	18.67	\pm	0.07	$	\\
57552.47619	&	$	18.56	\pm	0.03	$	\\
57553.48097	&	$	18.48	\pm	0.05	$	\\
57554.4719	&	$	18.56	\pm	0.03	$	\\
57555.4413	&	$	18.46	\pm	0.03	$	\\
57555.4413	&	$	18.45	\pm	0.03	$	\\
57555.46503	&	$	18.39	\pm	0.08	$	\\
57555.46503	&	$	18.45	\pm	0.02	$	\\
57564.44525	&	$	18.62	\pm	0.05	$	\\
57564.45274	&	$	18.63	\pm	0.05	$	\\
57567.48385	&	$	18.74	\pm	0.06	$	\\
57577.4716	&	$	18.77	\pm	0.04	$	\\
57578.4553	&	$	18.73	\pm	0.04	$	\\
57578.4553	&	$	18.52	\pm	0.04	$	\\
57578.48498	&	$	18.67	\pm	0.04	$	\\
57578.48498	&	$	18.69	\pm	0.04	$	\\
57584.45806	&	$	18.70	\pm	0.04	$	\\
57585.44955	&	$	18.69	\pm	0.10	$	\\
57585.44955	&	$	18.83	\pm	0.05	$	\\
57585.48817	&	$	18.80	\pm	0.05	$	\\
57586.43409	&	$	18.76	\pm	0.04	$	\\
57586.47677	&	$	18.84	\pm	0.04	$	\\
57596.45579	&	$	18.69	\pm	0.04	$	\\
57596.4886	&	$	18.77	\pm	0.05	$	\\
57599.46419	&	$	18.78	\pm	0.04	$	\\
57600.48294	&	$	18.64	\pm	0.03	$	\\
57601.47029	&	$	18.76	\pm	0.04	$	\\
57602.47299	&	$	18.83	\pm	0.05	$	\\
57603.47215	&	$	18.85	\pm	0.04	$	\\
57607.42403	&	$	18.75	\pm	0.04	$	\\
57607.4755	&	$	18.73	\pm	0.04	$	\\
57608.42244	&	$	18.72	\pm	0.04	$	\\
57608.46314	&	$	18.76	\pm	0.04	$	\\
57610.37778	&	$	18.90	\pm	0.05	$	\\
57610.40956	&	$	18.90	\pm	0.04	$	\\
57614.27914	&	$	18.52	\pm	0.04	$	\\
57614.31513	&	$	18.54	\pm	0.02	$	\\
57621.48494	&	$	18.77	\pm	0.08	$	\\
57625.51741	&	$	18.71	\pm	0.08	$	\\
57627.49786	&	$	18.87	\pm	0.05	$	\\
57627.50277	&	$	18.83	\pm	0.04	$	\\
57630.48997	&	$	18.80	\pm	0.04	$	\\
57637.50613	&	$	18.67	\pm	0.04	$	\\
57637.52178	&	$	18.79	\pm	0.08	$	\\
57643.49325	&	$	18.66	\pm	0.03	$	\\
57643.51429	&	$	18.68	\pm	0.03	$	\\
57655.28962	&	$	18.88	\pm	0.06	$	\\
57657.31213	&	$	18.71	\pm	0.04	$	\\
57660.32479	&	$	18.71	\pm	0.03	$	\\
57666.32325	&	$	18.86	\pm	0.05	$	\\
57673.20179	&	$	18.50	\pm	0.03	$	\\
57681.27201	&	$	18.87	\pm	0.05	$	\\
57734.08192	&	$	18.61	\pm	0.03	$	\\
57737.08603	&	$	18.63	\pm	0.06	$	\\
57737.10603	&	$	18.58	\pm	0.06	$	\\
57741.07583	&	$	18.45	\pm	0.04	$	\\
57741.091	&	$	18.52	\pm	0.05	$	\\
57798.12369	&	$   18.58	\pm	0.05	$	\\
\enddata
\end{deluxetable}

\begin{deluxetable}{cc}
\tablecaption{g-band Photometry}
\tablewidth{0pt}
\tablehead{Date(MJD) & Mag}
\startdata
55068.47442	&	$	19.46	\pm	0.03	$	\\
55182.10499	&	$	20.08	\pm	0.03	$	\\
55182.14066	&	$	20.18	\pm	0.04	$	\\
55182.17708	&	$	20.08	\pm	0.04	$	\\
55182.24007	&	$	20.25	\pm	0.09	$	\\
55182.27529	&	$	20.11	\pm	0.07	$	\\
55182.31123	&	$	19.95	\pm	0.07	$	\\
55182.34787	&	$	20.10	\pm	0.10	$	\\
56578.38672	&	$	20.16	\pm	0.04	$	\\
56578.42594	&	$	20.11	\pm	0.04	$	\\
56650.12354	&	$	20.07	\pm	0.10	$	\\
56655.10223	&	$	20.08	\pm	0.08	$	\\
56655.13466	&	$	20.09	\pm	0.06	$	\\
56657.09685	&	$	19.57	\pm	0.05	$	\\
56657.13672	&	$	19.45	\pm	0.03	$	\\
56907.24814	&	$	20.22	\pm	0.11	$	\\
56913.47892	&	$	20.30	\pm	0.09	$	\\
56913.51227	&	$	20.34	\pm	0.07	$	\\
56920.49168	&	$	18.17	\pm	0.07	$	\\
56920.5219	&	$	19.67	\pm	0.07	$	\\
56925.48542	&	$	20.32	\pm	0.08	$	\\
56925.51464	&	$	19.68	\pm	0.07	$	\\
56935.4322	&	$	20.17	\pm	0.03	$	\\
56936.38448	&	$	20.15	\pm	0.09	$	\\
56936.41448	&	$	20.52	\pm	0.11	$	\\
56940.39396	&	$	20.33	\pm	0.11	$	\\
56941.40683	&	$	19.97	\pm	0.10	$	\\
56944.36186	&	$	20.26	\pm	0.07	$	\\
56944.39101	&	$	20.30	\pm	0.07	$	\\
56946.37385	&	$	19.86	\pm	0.05	$	\\
56946.40318	&	$	20.24	\pm	0.08	$	\\
56949.35029	&	$	20.07	\pm	0.05	$	\\
56949.37965	&	$	20.07	\pm	0.04	$	\\
56950.34933	&	$	20.11	\pm	0.02	$	\\
56950.38148	&	$	19.92	\pm	0.13	$	\\
56951.36521	&	$	20.03	\pm	0.04	$	\\
56951.39894	&	$	19.90	\pm	0.06	$	\\
56952.35101	&	$	20.04	\pm	0.03	$	\\
56952.37968	&	$	20.14	\pm	0.04	$	\\
56953.34565	&	$	20.17	\pm	0.05	$	\\
56954.36273	&	$	20.31	\pm	0.05	$	\\
56954.39161	&	$	20.34	\pm	0.04	$	\\
56955.34692	&	$	20.32	\pm	0.06	$	\\
56955.37579	&	$	20.39	\pm	0.08	$	\\
56958.34337	&	$	20.32	\pm	0.05	$	\\
56965.194	&	$	20.42	\pm	0.37	$	\\
57255.32787	&	$	18.93	\pm	0.19	$	\\
57255.37765	&	$	18.81	\pm	0.10	$	\\
57256.3263	&	$	20.45	\pm	0.08	$	\\
57257.31646	&	$	20.30	\pm	0.10	$	\\
57257.35243	&	$	18.95	\pm	0.13	$	\\
57258.31311	&	$	19.11	\pm	0.11	$	\\
57270.40806	&	$	20.25	\pm	0.06	$	\\
57270.43794	&	$	20.45	\pm	0.14	$	\\
57271.40752	&	$	20.38	\pm	0.07	$	\\
57271.4372	&	$	20.59	\pm	0.07	$	\\
57286.28075	&	$	20.47	\pm	0.06	$	\\
57286.3111	&	$	20.36	\pm	0.03	$	\\
57655.44814	&	$	20.50	\pm	0.16	$	\\
57657.36159	&	$	20.20	\pm	0.06	$	\\
57660.37301	&	$	20.19	\pm	0.04	$	\\
57666.37122	&	$	20.44	\pm	0.06	$	\\
57673.16652	&	$	20.30	\pm	0.10	$	\\
57681.22238	&	$	20.24	\pm	0.12	$	\\
57763.10123	&	$	20.25	\pm	0.21	$	\\
\enddata
\end{deluxetable}


\listofchanges

\end{document}